\documentclass[oldversion]{aa} 
\usepackage{natbib}
\usepackage{amssymb}
\usepackage{graphicx}
\usepackage{txfonts}

\begin{document}
   \title{The evolution of the X-ray emission of HH~2}
   \subtitle{Investigating heating and cooling processes}

   \author{P. C. Schneider\inst{1}
          \and
          H. M. G\"unther\inst{2}
          \and
          J. H. M. M. Schmitt\inst{1}
          }

   \institute{Hamburger Sternwarte,
              Gojenbergsweg 112, 21029 Hamburg\\
              \email{cschneider@hs.uni-hamburg.de}
    \and
              Harvard-Smithsonian Center for Astrophysics, 60 Garden Street, Cambridge, MA, USA          
             }

   \date{Received .. / accepted ..}

  \abstract
   {
       Young stellar objects often drive powerful bipolar outflows which evolve on time scales of a few years. An increasing number of these outflows has been detected in X-rays implying the existence of million degree plasma almost co-spatial with the lower temperature gas observed in the optical and near-IR. The details of the heating and cooling processes of the X-ray emitting part of these so-called Herbig-Haro objects are still ambiguous, e.g., whether the cooling is dominated by expansion, radiation or thermal conduction.
       
       We present a second epoch {\it Chandra} observation of the first X-ray detected Herbig-Haro object (HH~2) and derive the proper-motion of the X-ray emitting plasma and its cooling history. We argue that the most likely explanation for the constancy of the X-ray luminosity, the alignment with the optical emission and the proper-motion is that the cooling is dominated by radiative losses leading to cooling times exceeding a decade. We explain that a strong shock caused by fast material ramming into slower gas in front of it about ten years ago can explain the X-ray emission while being compatible with the available multi-wavelength data of HH~2.
   }

   \maketitle
%

\section{Introduction}\vspace{-2mm}
During their evolution to the main-sequence young stellar objects (YSOs) can drive powerful outflows as long as accretion proceeds and it is now widely accepted that accretion and ejection are intimately connected \citep{Cabrit_1990, Hartigan_1995}. Protostellar jets can remove angular momentum from the system and thus allow accretion to proceed \citep{Matt_2008}.
The beautiful manifestations of the outflows are Herbig-Haro (HH) objects which are shocked regions within the outflow, often also called knots \citep[for an overview see ][]{Reipurth_2001}. 
These HH objects were originally discovered as nebular emission lines in star forming regions and this forbidden line emission is still a characteristic feature of HH objects. To date, HH objects have counter-parts from cm- to X-ray-wavelengths.
They evolve on time scales of a few years and thus multi-epoch observations strongly improve the constraints on the underlying physical processes such as the shock heating and the dominant cooling process.

Many observations of HH objects use optical forbidden emission lines (FELs) as tracers for the jet and the temperature range exceeding $T\gg 10^4$\,K is often neglected. However, this high energy part of the jet provides important clues on the highest velocity part of the jet and thus on the driving engine which has to be able to provide this high velocity material. The highest temperatures in protostellar jets have been investigated with X-ray observations tracing a plasma temperature $T\gtrsim10^6$\,K. 
HH~2 constitutes the first detection of X-ray emission from an HH object \citep{Pravdo_2001} and following this initial discovery, about a dozen protostellar jets have been found to emit X-rays. However, most of them are rather dim in X-rays and detailed studies of their properties are complicated and often ambiguous. Multi-epoch observations exist for two X-ray bright sources, the DG~Tau jet \citep[HH~158, ][]{Guedel_2005,Schneider_2008,Guedel_2008,Guedel_2011} and the L1551~IRS\,5 jet \citep[HH~154, ][]{Favata_2002,Bally_2003,Favata_2006,Schneider_2011}. They show stationary X-ray emission close to the driving sources and this behaviour greatly differs from what is observed in the optical, namely individual knots moving along the jet with proper-motion velocities of about 300\,km\,s$^{-1}$. However, there have also been claims of moving X-ray emission regions within protostellar jets \citep{Favata_2006, Stelzer_2009}. Furthermore, the stationary X-ray components are located much closer to their driving sources than most of the optically observed, moving knots \citep{Eisloeffel_1998} as a resolution of about $0\farcs1$ is needed to monitor the inner regions of protostellar jets, which are also often heavily absorbed. It is therefore a fundamental question whether the X-ray emission from HH objects at some distance to the driving source shows the same behavior as the close-in sources or whether there is a dichotomy with moving and stationary X-ray sources in protostellar outflows.

HH~2 is ideally suited to monitor the time evolution of its X-ray emission: It is well isolated so that no contamination from, e.g., its driving source is present, it is one of the brightest X-ray jets and the first observation dates back to 2000 so that the baseline is sufficiently long.

\subsection*{HH~1/2}
\begin{figure}
\centering
\fbox{\includegraphics[width=0.46\textwidth]{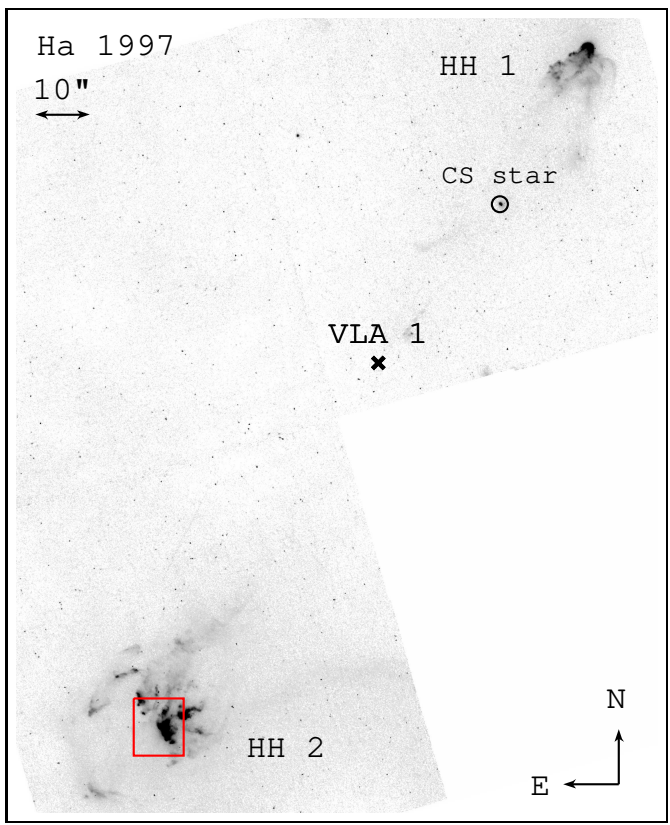}}
\caption{Overview of the HH~1/2 region (HST WFPC2 H$\alpha$ image from 1997). The red box indicates the region shown in Fig.~\ref{fig:ximage} and the Cohen-Schwartz star is also labeled (CS star). \label{fig:overview}}
\end{figure}

Together with HH~1, which is located at the opposite side of the deeply embedded driving source \citep[HH~1/2 VLA~1,][]{Pravdo_1985}, HH~2 ranks among the optically brightest HH objects. The overview of the HH~1/2 region in Fig.~\ref{fig:overview} shows that
HH~1 is north-west and HH~2 south-east of the central source which is invisible in the optical and near-IR. HH~1/2 VLA~1 is a class~0 source possessing an in-falling envelope \citep{Reipurth_1993}.  With its location in the Orion star forming region at a distance of 414\,pc \citep{Menten_2007}, multi-epoch HST observations \citep{Bally_2002, Hartigan_2011} revealed the evolution of the individual condensations (knots) within HH~1 and HH~2 over the course of only a few years. They show that the individual knots move with velocities of about 300\,km\,s$^{-1}$ almost in the plane of the sky. HH~2 appears as highly fragmented with some of the regions showing complex interaction between two knots or reverse facing shocks \citep{Hester_1998}. 
A bow-shock model for the optical/UV emission of HH~2 was presented by \citet{Hartigan_1987}. This model has a shock velocity of 160\,km\,s$^{-1}$, i.e., some 100\,km\,s$^{-1}$ lower than required by the existence of X-ray emission and a post-shock density of $n_e=2\times10^3$\,cm$^{-3}$.
A comprehensive summary of the work on HH~1/2 can be found in \citet{Raga_2011}.

Here, we present a second epoch {\it Chandra} observation of HH~2 with the aim to constrain the heating and the cooling mechanisms of X-ray emitting HH objects. 
Inspired by the optical data, we assume in the following that the evolution of the X-ray emission is rather slow, i.e., that the evolution of the X-ray emission of HH~2 is well behaved and not varying erratically so that we do not just catch it in two different states by chance.
\section{Observations and data reduction}

The new {\it Chandra} exposure uses the same ACIS-S chip as the archival observation from 2000, the details of the exposures can be found in Tab.~\ref{tab:Observations}.
We used CIAO~4.3 for the data analysis and reprocessed the observations to account for the VFAINT mode, removed any pixel randomization and reprojected the second epoch observation such that the brightest sources on the same ACIS chip as HH~2 are co-aligned. We estimate relative positional uncertainties of about $0\farcs1$ based on the remaining offsets of $0\farcs03$, $0\farcs08$ and $0\farcs14$\footnote{There is no remaining systematic trend.}.
Spectra and associated responses were extracted using the CIAO tool \texttt{specextract}.

To compare the X-ray data with the optical, specifically {\it Hubble} Space Telescope (HST) H$\alpha$ images of HH~2, we retrieved the corresponding data from the MAST archive. The optical data was already discussed in great depth by \citet{Hartigan_2011} and serves only as a morphological reference for the X-ray data here.
For the HST data, we relied on the pipeline processing. The alignment of the optical data was performed by aligning the Cohen-Schwartz star (V2669 Ori) and checking the rotational alignment with the HH objects themselves.

\begin{table}
\begin{minipage}[h]{0.49\textwidth}
\caption{Analyzed observations of HH~2.\label{tab:Observations}}
\centering
\setlength\tabcolsep{5pt}
\renewcommand\footnoterule{}
\begin{tabular}{l c c c c c}

\hline\hline
Date  & Detector & Filter & Obs./Prop. ID & exp. time\\
\hline
\multicolumn{5}{c}{X-ray ({\it Chandra}):}\\
2000-10-08 &  ACIS-S & - & 21 & 20\,ks\\
2010-12-06 &  ACIS-S & - & 12391 & 41\,ks\\
\hline
\multicolumn{5}{c}{H$\alpha$ (HST):}\\
1997-07-31 & WFPC2 &  F656N & 6794 & 2.0\,ks\\
2007-08-18 & WFPC2 & F656N & 11179 & 2.0\,ks\\
\hline
\end{tabular}
\end{minipage}
\end{table}

\section{Results}
Our new observation confirms the detection of HH~2 in X-rays; the new X-ray image as well as the archival one from 2001 are shown in Fig.~\ref{fig:ximage}. We find nine photons at the expected position of HH~2 using a circle with a radius of $2\farcs3$, while the expected number of background photons within this region is only 0.2 (0.3-1.0 keV) and 0.7 (0.3-3.0 keV), respectively. The exposure from 2000 collected 11 photons in a similarly shaped region on a background of 0.1 (0.3-1.0 keV) and 0.3 (0.3-3.0 keV) photons, respectively. The chance probability that any of the of two detections is a random background fluctuation is below $10^{-6}$.
In both observations, all photons attributable to HH~2 have energies below 1.0\,keV.
\subsection{The spatial photon distribution}
\begin{figure}
\centering
\fbox{\includegraphics[width=0.22\textwidth]{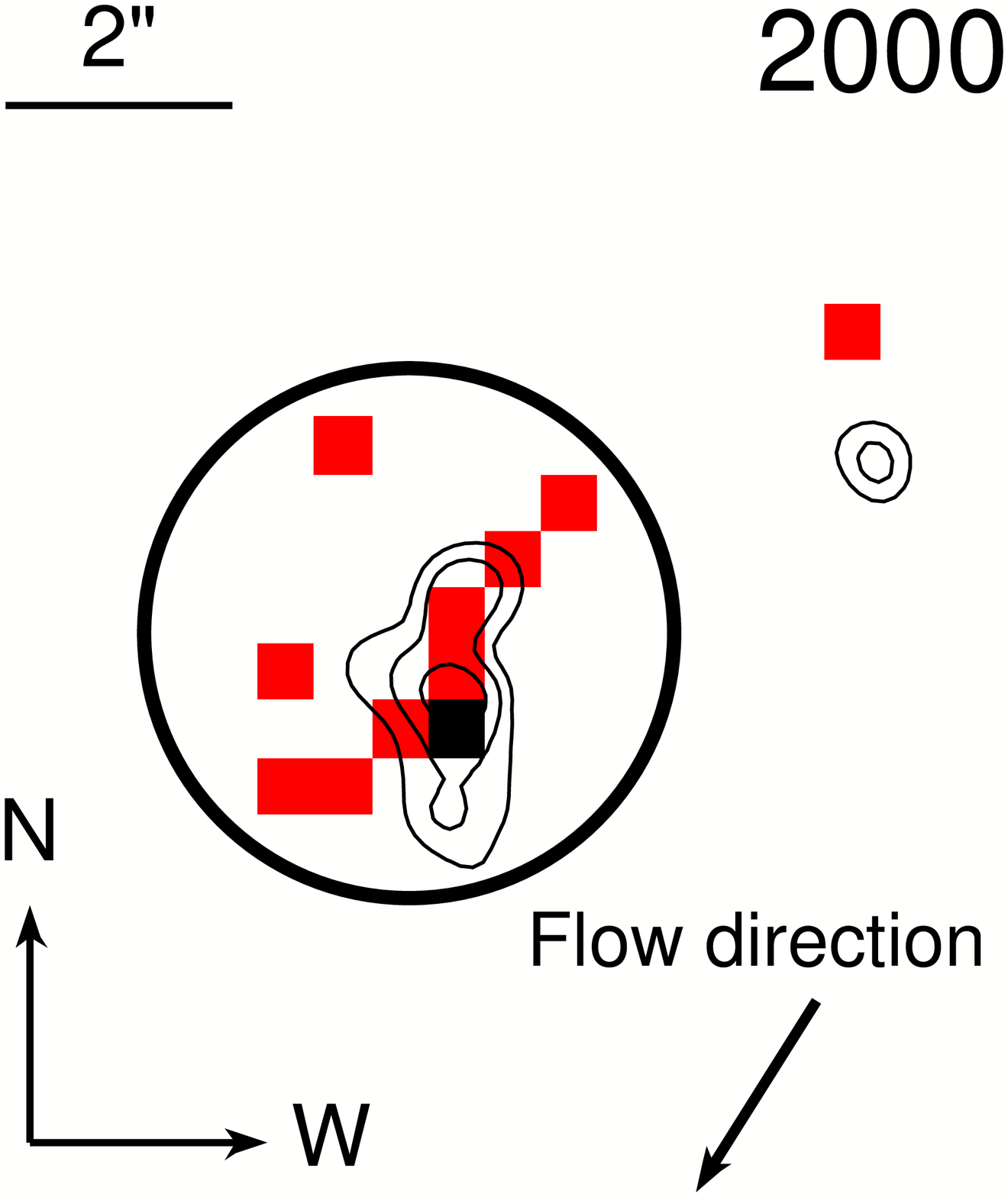}}
\fbox{\includegraphics[width=0.22\textwidth]{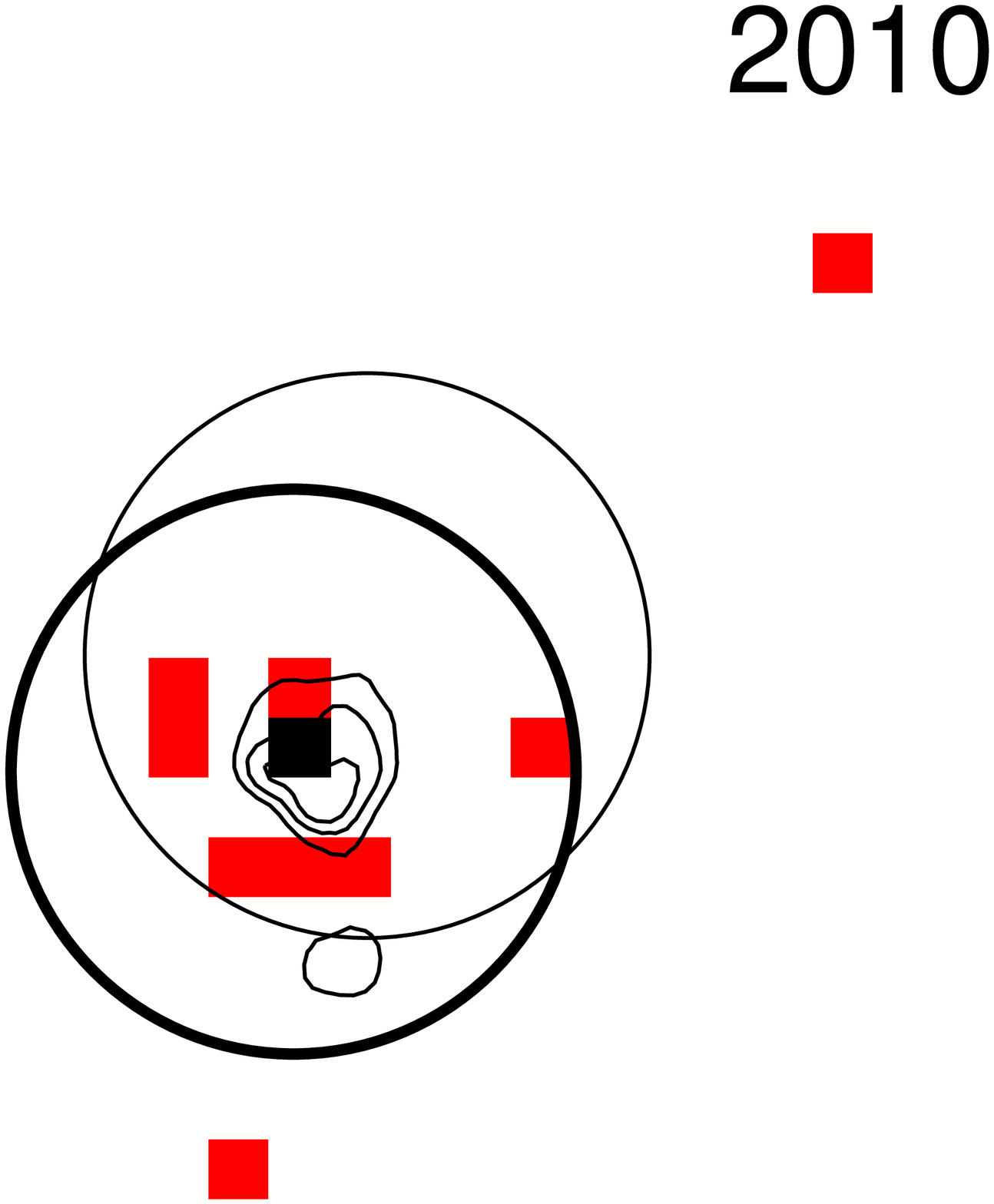}}
\caption{Proper-motion of the X-ray emitting plasma. Red squares denote pixel with single photon events and black squares display double photon events. \textbf{Left:} X-ray emission (0.3\,--1.0\,keV) from 2000 with H$\alpha$ contours from 1997 in black.  \textbf{Right:} X-ray emission from 2010 with H$\alpha$ contour from 2008. The thick circles of radius $2\farcs3$ coincide with the photon centroid, the thin circle indicates the position of the photons during the 2000 observation.\label{fig:ximage}}
\end{figure}

\begin{figure}
\centering
\includegraphics[width=0.49\textwidth]{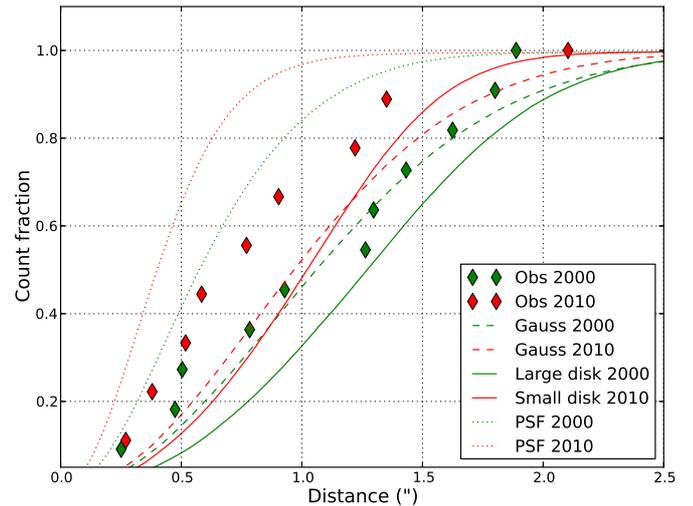}
\caption{Radial distribution of the photons from the centroid position with MARX models. The offaxis position of HH~2 affects the resulting photon distribution for the extended models only marginally (cf. Gauss 2000 and 2010) and the disk models are shown only for a single epoch.\label{fig:rDist}}
\end{figure}

We calculated the photon centroid 
using the photons within the estimated source region and find that the centroid position moved by $1\farcs1$ within the last decade.  The position angle of the offset is 148$^\circ$ and thus compatible with the optical jet axis.
The second moments of the spatial photon distributions
are $1\farcs20$ and $0\farcs85$ for the 2000 and 2010 exposures, resp.; the spatial profiles are shown in  Fig.~\ref{fig:rDist}.  Inclusion of the SE photon visible in Fig.~\ref{fig:ximage} for the 2010 observation changes the centroid only by $0\farcs3$ and the second moment increases to $1\farcs09$.
In order to address the significance of these results we used  MARX\footnote{\texttt{http://space.mit.edu/cxc/marx/}} simulations and two-sided KS-tests.

For our MARX simulations, we considered (a) a Gaussian source flux distribution and (b) a uniformly emitting disk. In both cases we used a spatial extent which reproduces the mean second moment of the two observations, i.e.,  $\sigma=1\farcs05$ for case (a) and a radius of $1\farcs4$ for case (b). Additionally, we investigated a uniformly emitting disk with $r=1\farcs7$ which we regard as an upper limit on the size of the emitting region (case c). For the Gaussian model (case a), the surface brightness drops below 10\,\% of the peak value at about $2\farcs3$; the circles shown in the Fig.~\ref{fig:ximage} have exactly this size and encircle $\gtrsim95$\% of the simulated photons.  
The simulations (a) and (b) show that it is highly unlikely to find the measured offset without any proper-motion ($<5$\%). The $68\%$ and 90\% errors of the relative offset are $0\farcs6$ and $0\farcs8$, respectively. The significance of proper-motion decreases to $88\,$\% for case (c) and, e.g., the 68\,\% error increases to $0\farcs7$.
Performing the simulations using the centroid calculated from photons within a radius of 4\arcsec{} increases the uncertainties by less than $0\farcs2$.

Comparing the differences of the observed second moments, the simulations indicate no significant difference in the underlying emitting region ($<80$\% sign.) which is compatible with the result of the two-sided KS test giving a probability of 77\,\% that both distributions are  not drawn from the same distribution. The impact of the off-axis position of HH~2 during the 2000 observation affects the result of the KS-test only marginally (the second moments change only by $\lesssim0\farcs1$).

In summary, we prefer the conservative interpretation concerning the evolution of the source shape that no significant \textit{expansion} occurred between the two exposures.
On the other hand, we regard it as very likely that the majority of the X-ray emitting plasma moved in the direction of the optical outflow.

In  Fig.~\ref{fig:ximage} the circles and the lowest contours give the boundary where the flux reaches 10\,\% of its maximum value. We note that the size scales of the optical and X-ray emission regions
appear to be similar, with the X-ray emission region being somewhat larger in both observations. A deconvolution of the optical images is not necessary here, as the spatial extent is much larger than the point spread function\footnote{The half energy size of HST is about $0\farcs1$ at these wavelengths.}.

\subsection{X-ray luminosity and energy distribution of the photons}
The observed X-ray count-rate of HH~2 dropped from $(3.9\pm1.8)\times10^{-4}$\,s$^{-1}$ to $(1.7\pm0.8)\times10^{-4}$\,s$^{-1}$ within the last decade. However, the effective area  at the very low energies of the photons also reduced by about a factor of two during that time so that a constant X-ray luminosity is likely. In order to investigate the properties of the plasma, we used XSPEC\,v12.5.0 \citep{XSPEC} to fit the spectra assuming an absorbed (\texttt{phabs}) optically thin single temperature plasma (\texttt{apec}) with half solar abundances. We used \texttt{cstatistic} and binned the spectra to 22\,eV bins. Due to the low number of photons in each individual observation, we decided to fix the absorption to different values as suggested by the literature, e.g, 
E(B-V)=0.34 \citep{Brugel_1981,BV_1982,Brugel_1982}, $A_V$=0.68 \citep{Boehm_1979} or E(B-V)=0.11 \citep{Hartmann_1984}.

We fitted the composite and the individual spectra, the results for the composite spectra are listed in Tab.~\ref{tab:Xresults}. The parameters for temperature and emission measure (EM) obtained by fitting the individual exposures are consistent with the results for the composite spectra showing that no significant spectral evolution occurred.  
The fitted temperature is $T\approx10^6$\,K while the value for the emission measure, and thus the unabsorbed X-ray luminosity, strongly depends on the assumed absorption (roughly $EM\sim10^{53}$\,cm$^{-3}$ and $\sim10^{29}\,$erg\,s$^{-1}$).
Inspection of the individual photons shows that most of the observed emission is consistent with being O~{\sc vii} emission with a small contribution of O~{\sc viii}. This further strengthens the results of the spectral fits as only for temperatures around or slightly above $10^6$\,K such a pattern is expected.
Although the slope of the effective area changed, the mean energy of the photons stays virtually constant with values of 543\,eV (0.3\,keV--3.0\,keV, 11 photons) and 532\,eV (9 photons). This can be explained with the fact that most of the detected photons originate from O~{\sc vii} emission ($E\approx561-574\,$eV), i.e., the majority of the photons is emitted within a small energy range.

In order to estimate which reduction in count-rate is still compatible with the data, we calculated the probability for detecting the observed photon numbers given two different expectation values, i.e., we calculated
\begin{equation}
P(n\leq 11,m\geq 9| \mu_1, \mu_2) = CDF(11,\mu_1) \times \left(1-CDF(8,\mu_2)\right)
\end{equation}
for different values of $\mu_1$ and $\mu_2$ using the Poissonian cumulative distribution function $CDF$. $CDF(x,\mu)$ gives the probability to find fewer or equal than $x$ photons given the expectation value $\mu$. Requiring $P>0.32\, (1\sigma)$ the ratio $\mu_1 / \mu_2$ is less than 1.2 while requiring $P>0.1$ this ratio increases to 1.8.
Therefore, the maximum decrease in X-ray luminosity of HH~2 is a factor 1.2 ($1\sigma$) or 1.8 (90\,\% conf.) assuming no spectral evolution as indicated above.

\begin{figure}
\centering
\includegraphics[width=0.46\textwidth]{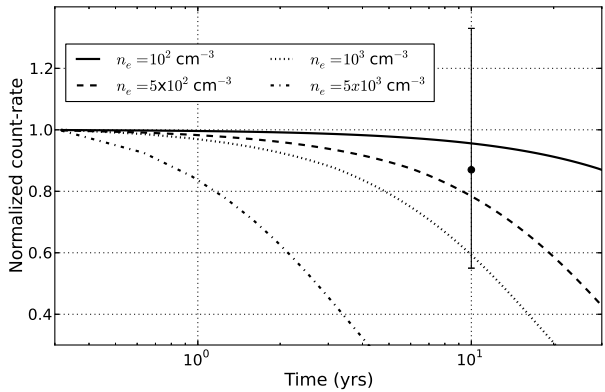}
\caption{Evolution of the expected count rate assuming $N_H=1.2\times10^{21}\,$cm$^{-2}$. The data point shows the measured count rate evolution corrected for the changes in the effective area.  \label{fig:cooling}}
\end{figure}

\section{Discussion}
Both observations are separated by one decade which is sufficient to exclude some scenarios for the heating and cooling of the X-ray emitting plasma based on the spatial evolution of the X-ray emitting region and the evolution of the X-ray luminosity.

\subsection{Proper-motion of the X-ray emission region and its spatial morphology}
The new data show proper-motion of the million degree plasma in the direction of the optical knot with a nominal velocity of $v_{X-ray}\approx210\,$km\,s$^{-1}$. The data exclude both, a stationary source and velocities exceeding 380\,km\,s$^{-1}$, on the 90\,\% confidence level. The nominal centroid proper-motion is lower than the velocities derived for individual regions within HH~2 knot~H from the optical data \citep[$\sim$300\,km\,s$^{-1}$, ][]{Bally_2002} but considering the relatively large error on the X-ray proper-motion, a velocity as high as that of the H$\alpha$ emitting plasma is possible within the $1\sigma$ error. 
This result for the proper-motion of the X-ray emission region of HH~2 represents the most accurate measurement of proper-motion of X-ray emitting HH objects to date as there is no ambiguity in assigning the initial position of the X-ray emission region and there is no contamination from other X-ray sources, e.g., from the driving source. The major uncertainties for HH~2 are the structure of the emission region and the low photon numbers. 
Longer exposures are required to overcome these issues and would also reveal how the X-ray and H$\alpha$ emission  regions differ. 

The derived velocity agrees very well with the  velocity of $\approx200$\,km\,s$^{-1}$ of the outer X-ray emitting jet of DG~Tau \citep[$T\approx3.4\times10^6$\,K, ][]{Guedel_2011}. On the other hand, the velocity of HH~2 is considerably smaller than the X-ray emitting knot in the Z~CMa jet \citep{Stelzer_2009}. However, the initial position of the latter knot is rather uncertain and the estimated X-ray proper-motion exceeds the optical estimate in that case (contrary to HH~2 and DG~Tau). 
We do not include the jet of L\,1551~IRS\,5 here because the majority of the X-ray emission from this jet is in fact stationary and no reliable proper-motion value can be derived \citep{Schneider_2011}. 
Considering the temperatures of the  X-ray plasmas, it appears that higher post-shock proper-motion leads to higher plasma temperatures. A similar trend has been found for individual knots in HH~2 between the proper-motion velocity and the line width which is an indicator of the actual shock velocity by \citet{Eisloeffel_1998}.  However, we refrain from claiming a trend because the number of sources is low and because the post-shock velocity is not necessarily related to the actual shock-velocity.

\begin{table}
\begin{minipage}[h]{0.49\textwidth}
\caption{Results of the specral fits to the X-ray data.\label{tab:Xresults}}
\centering
\setlength\tabcolsep{2.5pt}
\renewcommand\footnoterule{}
\begin{tabular}{l c c c c c}
\hline
\hline
& $A_V=0.34$ & $A_V=0.68$ & $A_V=1.05$ & $A_V$: free \\
Parameter \\
\hline\\
$N_H$ (cm$^{-2}$) & $6\times10^{20}$ & $1.2\times10^{21}$ & $1.9\times10^{21}$ & $4.4_{-4.4}^{+8.6}\times10^{21}$\\[0.1cm]
$kT$ (keV) & $0.11\pm0.02$ & $0.10^{+0.02}_{-0.01}$ & $0.09\pm0.02$ & $0.06^{+0.1}_{-0.03}$\\[0.1cm]
$EM$ ($10^{52}$\,cm$^{-3}$) & $2.4^{+2.7}_{-1.0}$ & $7.0^{+7.5}_{-7.0}$ & $23.4^{+62.1}_{-22.8}$ & $1842.1_{-1842.1}^{+9261.6}$ \\[0.1cm]
Luminosity  & 0.9 & 1.8 & 4.3 & 116.3\\[0.1cm]
\hspace*{0.3cm}($10^{29}\,$erg\,s$^{-1}$)\\
\hline
\end{tabular}
\end{minipage}
\end{table}

\subsection{Heating and cooling}
In principle, there are two possibilities for the constant X-ray luminosity: Either the cooling time is rather long or the X-ray emitting plasma is constantly replenished. By investigating the importance of adiabatic, radiative and conductive cooling we argue that slow cooling is the most likely scenario and discuss the resulting implications for the heating process.

\subsubsection{Cooling}
The X-ray emission region of HH~2 is incompatible with a point source in both exposures, as already shown from the first epoch data by \citet{Pravdo_2001}. Furthermore, we do not find any evidence for an increasing (projected) size of the emission region. 
Assuming a sphere with the same radius as the circle in Fig.~\ref{fig:ximage} and a volume filling factor of unity, we derive electron densities of $n_e\sim10^2-10^3$\,cm$^{-3}$ depending on the exact value for the emission measure. For smaller emission regions, i.e., if the emission is composed of individual clumps, the electron density will increase. 
With these densities the X-ray emitting material can be in thermal pressure equilibrium with the other temperature components and the absence of significant source variability makes cooling by expansion unlikely.

Using the CHIANTI atomic database \citep{Dere_1997,Dere_2009}, we estimate that the observed count rate would drop by a factor of 1.4 within 10\,years by radiative cooling assuming $n_e=5\times10^2$\,cm$^{-3}$ or by 20\% within 30\,years assuming $n_e=100$\,cm$^{-3}$ (ignoring any further drop in the effective area, see Fig.~\ref{fig:cooling}), which is perfectly compatible with the data in the absence of reheating. The variation in the temperature is even smaller: For $n_e=10^3\,$cm$^{-3}$ the temperature changes by only 10\,\% within a decade so that the appearance of a constant plasma temperature is fully compatible with radiative cooling.

Using the approximation given in \citet{Orlando_2005} for the conductive cooling time of a million degree plasma embedded in a gas of much lower temperature, we find that the conductive cooling time exceeds $10^4$\,years in the case of HH~2 and is thus negligible compared to radiative cooling. 

In summary, the dominant cooling process is most likely radiation as the cooling time of thermal conduction is too long, while the data does not indicate cooling by expansion. 

\subsubsection{Heating}
Currently, knot~H is in a phase of being overtaken by the faster knot M \citep{Eisloeffel_1994,Hartigan_2011}, so that the X-ray emission might  not be intrinsic to knot~H but rather to the interaction of both knots.  The measured velocities for the two knots were 220 and 490\,km\,s$^{-1}$ \citep{Eisloeffel_1994} resulting in a shock velocity of $v_{sh}=270$\,km\,s$^{-1}$. The derived temperature of the X-ray emitting plasma of about $T\approx10^6$\,K requires shock velocities of 260\,km\,s$^{-1}$ \citep{Raga_2002}, i.e., the expected shock velocity is precisely the value needed for the observed X-ray emitting plasma. The expected post-shock proper-motion velocity would be 290\,km\,s$^{-1}$ which is compatible with the X-ray and H$\alpha$ data. Even the nominal differences between the proper-motion of X-ray and H$\alpha$ emission would  not lead to detectable offsets during this time span. This picture can also explain the non-detection of O~{\sc vi} emission (peak formation temperature $3.2\times10^5\,$K) in the HUT spectrum taken in 1995 \citep{Raymond_1997} by assuming that knot~M did not encounter knot~H by that time and the existence of X-ray emission in 2001 would be a fortunate event in the history of HH~2.

Concerning the properties of the outflow, this scenario requires a jet with vastly different velocity components which move approximately into the same direction. As the flow-times of the two ejecta to the current position differ, both knots result from different outflow epochs. However, it is not clear whether knot~H is regularly fed by a high-velocity, pulsed outflow or whether these events are rather stochastic and less frequent. Since knot~M was observable before encountering knot~H, it also experienced shocks before. These shocks, however, have to be of much smaller shock velocity, first, because knot~M preserved its high velocity and, second, because no O~{\sc vi} emission was detected from the corresponding region in the 1995 HUT observation \citep{Raymond_1997}.

The second possibility to explain the constant X-ray luminosity and temperature is to assume that the million degree plasma is constantly refueled by a number of smaller shocks approximately uniformly distributed over the time.
Although the interaction between knot~M and knot~H persists for some time, the dramatic changes of the H$\alpha$ emission  \citep{Hartigan_2011} argue against a quasi-static shock scenario for the X-ray emission. Furthermore, the data does not indicate  higher photon energies closer to the centroid position as one would expect if the X-ray emitting plasma is constantly replenished at the position of the H$\alpha$ producing shock followed by expansion and cooling of the post-shock plasma. 

The available X-ray data of HH~2 therefore suggest that a single strong shock about a decade ago caused the heating to X-ray emitting temperatures since constant re-heating is not required, i.e., it is very  likely that individual strong shocks heat a fraction of the material to X-ray emitting temperatures.

\section{Summary and conclusion}
As protostellar jets and consequently also HH objects are dynamic objects evolving on time scales of years, our second epoch observation of HH~2 allows to further constrain their heating and cooling processes. The new data show that the X-ray emission of HH~2 moves into the same direction as the optical knots. The X-ray proper-motion velocity is  lower but still compatible to that of the H$\alpha$ emission. The constant X-ray luminosity, the absence of strong expansion and the unimportance of conductive cooling imply that the cooling of HH~2 is dominated by radiation with a rather long radiative cooling time of a few decades. 

With these facts, the most likely scenario is that the X-ray emitting part of the knot results from a single, strong shock when a bullet of high speed material rams into much slower material in front of it. For HH~2, these two components are probably the high velocity knot~M and the slower knot~H since their velocity difference is exactly the shock velocity needed for the observed X-ray emission. There is no evidence for a continuous or significant replenishment of hot, X-ray emitting plasma but a significant fraction of the optically observed gas might result from additional slower shocks. This naturally explains that both components are (almost) co-spatial and move with comparable velocities into the same direction.

In order to measure any separation between the X-ray and H$\alpha$ emitting part, one needs longer exposures  resolving the structure of the X-ray emission. 
Future observations will also reveal the true X-ray cooling time which we could not address with our second epoch observation due to the lack of significant cooling within the last decade.

\begin{acknowledgements}
PCS was supported by the DLR under grant 50 OR 1112.
HMG was supported by NASA through Chandra Award Number GO1-12067X issued by the Chandra X-ray Observatory Center, which is operated by the Smithsonian Astrophysical Observatory for and on behalf of NASA under contract NAS8-03060.
The work used data obtained by the Chandra X-ray Observatory and by the Hubble Space Telescope. 
\end{acknowledgements}

\bibliographystyle{aa}
\bibliography{main}

\begin{thebibliography}{35}
\expandafter\ifx\csname natexlab\endcsname\relax\def\natexlab#1{#1}\fi

\bibitem[{{Arnaud}(1996)}]{XSPEC}
{Arnaud}, K.~A. 1996, in Astronomical Society of the Pacific Conference Series,
  Vol. 101, Astronomical Data Analysis Software and Systems V, ed. G.~H.
  {Jacoby} \& J.~{Barnes}, 17--+

\bibitem[{{Bally} {et~al.}(2003){Bally}, {Feigelson}, \&
  {Reipurth}}]{Bally_2003}
{Bally}, J., {Feigelson}, E., \& {Reipurth}, B. 2003, \apj, 584, 843

\bibitem[{{Bally} {et~al.}(2002){Bally}, {Heathcote}, {Reipurth}, {Morse},
  {Hartigan}, \& {Schwartz}}]{Bally_2002}
{Bally}, J., {Heathcote}, S., {Reipurth}, B., {et~al.} 2002, \aj, 123, 2627

\bibitem[{{Boehm} \& {Brugel}(1979)}]{Boehm_1979}
{Boehm}, K.~H. \& {Brugel}, E.~W. 1979, \aap, 74, 297

\bibitem[{{Boehm-Vitense} {et~al.}(1982){Boehm-Vitense}, {Cardelli}, {Nemec},
  \& {Boehm}}]{BV_1982}
{Boehm-Vitense}, E., {Cardelli}, J.~A., {Nemec}, J.~M., \& {Boehm}, K.~H. 1982,
  \apj, 262, 224

\bibitem[{{Brugel} {et~al.}(1981){Brugel}, {Boehm}, \& {Mannery}}]{Brugel_1981}
{Brugel}, E.~W., {Boehm}, K.~H., \& {Mannery}, E. 1981, \apj, 243, 874

\bibitem[{{Brugel} {et~al.}(1982){Brugel}, {Seab}, \& {Shull}}]{Brugel_1982}
{Brugel}, E.~W., {Seab}, C.~G., \& {Shull}, J.~M. 1982, \apjl, 262, L35

\bibitem[{{Cabrit} {et~al.}(1990){Cabrit}, {Edwards}, {Strom}, \&
  {Strom}}]{Cabrit_1990}
{Cabrit}, S., {Edwards}, S., {Strom}, S.~E., \& {Strom}, K.~M. 1990, \apj, 354,
  687

\bibitem[{{Dere} {et~al.}(1997){Dere}, {Landi}, {Mason}, {Monsignori Fossi}, \&
  {Young}}]{Dere_1997}
{Dere}, K.~P., {Landi}, E., {Mason}, H.~E., {Monsignori Fossi}, B.~C., \&
  {Young}, P.~R. 1997, \aaps, 125, 149

\bibitem[{{Dere} {et~al.}(2009){Dere}, {Landi}, {Young}, {Del Zanna},
  {Landini}, \& {Mason}}]{Dere_2009}
{Dere}, K.~P., {Landi}, E., {Young}, P.~R., {et~al.} 2009, \aap, 498, 915

\bibitem[{{Eisl{\"o}ffel} \& {Mundt}(1998)}]{Eisloeffel_1998}
{Eisl{\"o}ffel}, J. \& {Mundt}, R. 1998, \aj, 115, 1554

\bibitem[{{Eisl{\"o}ffel} {et~al.}(1994){Eisl{\"o}ffel}, {Mundt}, \&
  {Bohm}}]{Eisloeffel_1994}
{Eisl{\"o}ffel}, J., {Mundt}, R., \& {Bohm}, K.-H. 1994, \aj, 108, 1042

\bibitem[{{Favata} {et~al.}(2006){Favata}, {Bonito}, {Micela}, {Fridlund},
  {Orlando}, {Sciortino}, \& {Peres}}]{Favata_2006}
{Favata}, F., {Bonito}, R., {Micela}, G., {et~al.} 2006, \aap, 450, L17

\bibitem[{{Favata} {et~al.}(2002){Favata}, {Fridlund}, {Micela}, {Sciortino},
  \& {Kaas}}]{Favata_2002}
{Favata}, F., {Fridlund}, C.~V.~M., {Micela}, G., {Sciortino}, S., \& {Kaas},
  A.~A. 2002, \aap, 386, 204

\bibitem[{{G\"udel} {et~al.}(2011){G\"udel}, {Audard}, {Bacciotti}, {Bary},
  {Briggs}, {Cabrit}, {Carmona}, {Codella}, {Dougados}, {Eisloeffel}, {Gueth},
  {Guenther}, {Herczeg}, {Kundurthy}, {Matt}, {Mutel}, {Ray}, {Schmitt},
  {Schneider}, {Skinner}, \& {van Boekel}}]{Guedel_2011}
{G\"udel}, M., {Audard}, M., {Bacciotti}, F., {et~al.} 2011, ArXiv e-prints

\bibitem[{{G{\"u}del} {et~al.}(2008){G{\"u}del}, {Skinner}, {Audard}, {Briggs},
  \& {Cabrit}}]{Guedel_2008}
{G{\"u}del}, M., {Skinner}, S.~L., {Audard}, M., {Briggs}, K.~R., \& {Cabrit},
  S. 2008, \aap, 478, 797

\bibitem[{{G{\"u}del} {et~al.}(2005){G{\"u}del}, {Skinner}, {Briggs}, {Audard},
  {Arzner}, \& {Telleschi}}]{Guedel_2005}
{G{\"u}del}, M., {Skinner}, S.~L., {Briggs}, K.~R., {et~al.} 2005, \apjl, 626,
  L53

\bibitem[{{Hartigan} {et~al.}(1995){Hartigan}, {Edwards}, \&
  {Ghandour}}]{Hartigan_1995}
{Hartigan}, P., {Edwards}, S., \& {Ghandour}, L. 1995, \apj, 452, 736

\bibitem[{{Hartigan} {et~al.}(2011){Hartigan}, {Frank}, {Foster}, {Wilde},
  {Douglas}, {Rosen}, {Coker}, {Blue}, \& {Hansen}}]{Hartigan_2011}
{Hartigan}, P., {Frank}, A., {Foster}, J.~M., {et~al.} 2011, \apj, 736, 29

\bibitem[{{Hartigan} {et~al.}(1987){Hartigan}, {Raymond}, \&
  {Hartmann}}]{Hartigan_1987}
{Hartigan}, P., {Raymond}, J., \& {Hartmann}, L. 1987, \apj, 316, 323

\bibitem[{{Hartmann} \& {Raymond}(1984)}]{Hartmann_1984}
{Hartmann}, L. \& {Raymond}, J.~C. 1984, \apj, 276, 560

\bibitem[{{Hester} {et~al.}(1998){Hester}, {Stapelfeldt}, \&
  {Scowen}}]{Hester_1998}
{Hester}, J.~J., {Stapelfeldt}, K.~R., \& {Scowen}, P.~A. 1998, \aj, 116, 372

\bibitem[{{Matt} \& {Pudritz}(2008)}]{Matt_2008}
{Matt}, S. \& {Pudritz}, R. 2008, in Astronomical Society of the Pacific
  Conference Series, Vol. 384, 14th Cambridge Workshop on Cool Stars, Stellar
  Systems, and the Sun, ed. {G.~van Belle}, 339--+

\bibitem[{{Menten} {et~al.}(2007){Menten}, {Reid}, {Forbrich}, \&
  {Brunthaler}}]{Menten_2007}
{Menten}, K.~M., {Reid}, M.~J., {Forbrich}, J., \& {Brunthaler}, A. 2007, \aap,
  474, 515

\bibitem[{{Orlando} {et~al.}(2005){Orlando}, {Peres}, {Reale}, {Bocchino},
  {Rosner}, {Plewa}, \& {Siegel}}]{Orlando_2005}
{Orlando}, S., {Peres}, G., {Reale}, F., {et~al.} 2005, \aap, 444, 505

\bibitem[{{Pravdo} {et~al.}(2001){Pravdo}, {Feigelson}, {Garmire}, {Maeda},
  {Tsuboi}, \& {Bally}}]{Pravdo_2001}
{Pravdo}, S.~H., {Feigelson}, E.~D., {Garmire}, G., {et~al.} 2001, \nat, 413,
  708

\bibitem[{{Pravdo} {et~al.}(1985){Pravdo}, {Rodriguez}, {Curiel}, {Canto},
  {Torrelles}, {Becker}, \& {Sellgren}}]{Pravdo_1985}
{Pravdo}, S.~H., {Rodriguez}, L.~F., {Curiel}, S., {et~al.} 1985, \apjl, 293,
  L35

\bibitem[{{Raga} {et~al.}(2002){Raga}, {Noriega-Crespo}, \&
  {Vel{\'a}zquez}}]{Raga_2002}
{Raga}, A.~C., {Noriega-Crespo}, A., \& {Vel{\'a}zquez}, P.~F. 2002, \apjl,
  576, L149

\bibitem[{{Raga} {et~al.}(2011){Raga}, {Reipurth}, {Cant{\'o}},
  {Sierra-Flores}, \& {Guzm{\'a}n}}]{Raga_2011}
{Raga}, A.~C., {Reipurth}, B., {Cant{\'o}}, J., {Sierra-Flores}, M.~M., \&
  {Guzm{\'a}n}, M.~V. 2011, \rmxaa, 47, 425

\bibitem[{{Raymond} {et~al.}(1997){Raymond}, {Blair}, \& {Long}}]{Raymond_1997}
{Raymond}, J.~C., {Blair}, W.~P., \& {Long}, K.~S. 1997, \apj, 489, 314

\bibitem[{{Reipurth} \& {Bally}(2001)}]{Reipurth_2001}
{Reipurth}, B. \& {Bally}, J. 2001, \araa, 39, 403

\bibitem[{{Reipurth} {et~al.}(1993){Reipurth}, {Heathcote}, {Roth},
  {Noriega-Crespo}, \& {Raga}}]{Reipurth_1993}
{Reipurth}, B., {Heathcote}, S., {Roth}, M., {Noriega-Crespo}, A., \& {Raga},
  A.~C. 1993, \apjl, 408, L49

\bibitem[{{Schneider} {et~al.}(2011){Schneider}, {G{\"u}nther}, \&
  {Schmitt}}]{Schneider_2011}
{Schneider}, P.~C., {G{\"u}nther}, H.~M., \& {Schmitt}, J.~H.~M.~M. 2011, \aap,
  530, A123

\bibitem[{{Schneider} \& {Schmitt}(2008)}]{Schneider_2008}
{Schneider}, P.~C. \& {Schmitt}, J.~H.~M.~M. 2008, \aap, 488, L13

\bibitem[{{Stelzer} {et~al.}(2009){Stelzer}, {Hubrig}, {Orlando}, {Micela},
  {Mikul{\'a}{\v s}ek}, \& {Sch{\"o}ller}}]{Stelzer_2009}
{Stelzer}, B., {Hubrig}, S., {Orlando}, S., {et~al.} 2009, \aap, 499, 529

\end{thebibliography}

\end{document}